# Suppression of excess oxygen for environmentally stable amorphous In-Si-O thin-film transistors

Shinya Aikawa,[1,2, a] Nobuhiko Mitoma,[1] Takio Kizu,[1] Toshihide Nabatame,[3] and Kazuhito Tsukagoshi[1,a]

[1]International Center for Materials Nanoarchitectonics (WPI-MANA), National Institute for Materials Science (NIMS), Tsukuba, Ibaraki 305-0044, Japan

[2]Research Institute for Science and Technology, Kogakuin University, Hachioji, Tokyo 192-0015, Japan

[3]MANA Foundry and MANA Advanced Device Materials Group, National Institute for Materials Science (NIMS), Tsukuba, Ibaraki 305-0044, Japan

___________________________

[a] Author to whom correspondence should be addressed. E-mail addresses: aikawa@cc.kogakuin.ac.jp and TSUKAGOSHI.Kazuhito@nims.go.jp

**We discuss the environmental instability of amorphous indium oxide ($InO_x$)-based thin-film transistors (TFTs) in terms of the excess oxygen in the semiconductor films. A comparison between amorphous $InO_x$ doped with low and high concentrations of oxygen binder ($SiO_2$) showed that out-diffusion of oxygen molecules causes drastic changes in the film conductivity and TFT turn-on voltages. Incorporation of sufficient $SiO_2$ could suppress fluctuations in excess oxygen because of the high oxygen bond-dissociation energy and low Gibbs free energy. Consequently, the TFT operation became rather stable. The results would be useful for the design of reliable oxide TFTs with stable electrical properties.**

Amorphous oxide semiconductors have been widely studied for use in semiconductor devices or transparent conductive films owing to their unique carrier transport properties.[1,2] Indium oxide (InO$_x$)-based semiconductors are particularly attractive for use as switching elements in thin-film transistors (TFTs) for high-definition flat panel displays. This is because InO$_x$-based semiconductors have a high electron mobility originating from the direct overlap of the isotropic *s* orbitals of In atoms.[3] Also, the fabrication process is compatible with conventional amorphous Si TFTs.[2] However, it is generally difficult to control the formation of oxygen vacancies (V$_O$), which generate free carriers in amorphous InO$_x$-based semiconductors.[4,5] When oxygen molecules are desorbed from an oxide semiconductor because of changes in the atmospheric environment, the TFT properties are affected.[6-8] This is a serious issue for processes that occur in a reducing atmosphere[9] such as SiN$_x$ passivation[10] and dry etching of the source/drain electrodes.[11] To stabilize the TFT characteristics by controlling the creation of V$_O$, an oxygen binding dopant is necessary. This dopant should be able to incorporate oxygen atoms inside the film during sputtering and cannot easily dissociate the bonded oxygen. Many oxide materials have been considered as dopant candidates so far.[12] For example, In–W–O,[13,14] In–Si–O,[15-17] In–Si–W–O,[18] In–C–Zn–O,[19] In–Ga–Zn–O,[20,21] In–Si–Zn–O,[22,23] and In–Ta–Zn–O[24,25] have been developed for InO$_x$-based active TFT channels.

We have previously developed amorphous $InO_x$-based semiconductors by focusing on materials that have a higher oxygen bond-dissociation energy (BDE) to suppress oxygen diffusion out of the film.[15] With increasing BDE, changes in the film conductivity are less sensitive to the oxygen partial pressure ($P_{O2}$) during sputtering, thus a wide range of sputtering conditions can be obtained when $SiO_2$ is doped into the $InO_x$ host as an oxygen binder.[15] Here we present the influence of the storage environment on the electrical stability of the oxide thin-film deposited at higher $P_{O2}$. Even though $SiO_2$ exhibits a high BDE (799.6 kJ/mol)[26] and acts as an effective $V_O$ suppressor, the resulting TFTs with a low concentration of oxygen binding dopant changes significantly from displaying semiconductor to metallic properties after being stored under vacuum. To solve this issue, we fabricated TFTs using an In–Si–O film with a higher dopant concentration. The oxide film would remain a semiconductor even though the films were prepared at a lower $P_{O2}$. This is because $SiO_2$ with a high BDE enables effective incorporation of oxygen atoms into the film. Using the highly doped In–Si–O film, degradation of TFT properties caused by oxygen out-diffusion is prevented and the long-term environmental stability is dramatically improved. This is very important not only for long term operational reliability but for efficient mass production of oxide TFTs.

The TFTs were fabricated on a heavily *p*-doped Si substrate with 250 nm of thermally grown $SiO_2$. The semiconductor film was deposited using DC magnetron sputtering of indium oxide doped with a low SiO$_2$ concentration (In$_2$O$_3$:SiO$_2$ = 97:3 by weight), hereafter denoted as ISO3. Using sputtering gas of Ar mixed with O$_2$ (1:1 ratio) at 0.25 Pa total pressure, the 10-nm-thick ISO3 active channels were formed through a stencil shadow mask. The sputtering power used to deposit films was 200 W. Prior to deposition of Mo source/drain electrodes (thickness: 40 nm) by electron-beam evaporation, the ISO3 film was thermally annealed at 250 °C for 30 min in air. After formation of the electrodes, the TFTs were annealed in air at 150 °C for 30 min and then annealed again in O$_3$ at the same temperature for 5 min. No passivation layers were formed on the channels. Electrical characterization was performed using a semiconductor parameter analyzer (Agilent 4156C) in air and in the dark. The fabricated TFTs were sequentially measured in different three states: just after the fabrication (as-fabricated); after storage in a vacuum desiccator (~10 Pa) for 3 months after the first measurement (stored in vacuum); after exposure to ambient air for 2 weeks after the second measurement (exposure to air).

Figure 1 shows *I-V* characteristics of the TFTs with ISO3 films. For the as-fabricated TFTs, transfer characteristics with different channel lengths (50−350 μm in 50-μm steps) and output characteristics with a 50-μm channel length are presented in Fig. 1(a) and (d), respectively. The maximum drain current ($I_D$) increased with decreasing channel length ($L_{ch}$). The turn-on voltage ($V_{ON}$), which is defined as the gate voltage ($V_{GS}$) at which the $I_D$ begins to increase, was almost constant. After being stored in a vacuum desiccator, however, the $V_{ON}$ was significantly shifted to more negative voltages in the shorter channel TFTs. Even when $V_{GS} = -100$ V was applied, $I_D$ was not fully turned off in the TFT with a 50-μm channel length (Fig. 1(b) and (e)). After being exposed to air for two weeks, the $V_{ON}$ in each channel length tended to return to zero (Fig. 1(c) and (f)).

To observe possible changes in the TFT properties after vacuum storage and air exposure, we extracted the contact resistance ($R_c$) and channel resistivity ($r_{ch}$) of the TFTs using the transfer line method (TLM). As shown in Fig. 2(a), the width-normalized total resistance ($R_{total}W$) extracted at a drain voltage ($V_{DS}$) of 1 V as a function of $L_{ch}$ indicates that the as-fabricated and exposure to air TFTs fit the linear relation, which is expressed by[17,27]

$$R_{total}W = r_{ch}L_{ch}W + R_cW, \quad (1)$$

where $W$ is the channel width, which was fixed at 1000 μm in this study. The $R_c$ takes into account the contact resistance at the source and drain electrodes ($R_c = R_{source} + R_{drain}$). The $L_{ch}$ was determined from optical microscope images. For the TFTs stored in vacuum, the longer channels (> 200 μm) showed a nonlinear $I_D$ relation near zero $V_{DS}$. The dotted lines in the middle of Fig. 2(a) were phenomenologically fitted using $R_{total}W = r_{ch}L_{ch}W\exp(L_{ch}/L_0) + R_cW$, where $L_0$ is the localization length. In the long channel regime, carriers are strongly localized in a disordered system[28] because of high density defects originating from the high $V_O$ density.

The nonlinear behavior at long length scales has been observed in other materials.[29-31] The localization effect of the ISO3 film will be discussed in more detail elsewhere.

Here, an intrinsic field-effect mobility ($\mu_{FE\text{-}i}$) is compared among the three cases using the following equation:[32]

$$1/r_{ch} = C_i W \mu_{FE-i}(V_{GS} - V_{th-i}), \qquad (2)$$

where $V_{th\text{-}i}$ is the intrinsic threshold voltage and $C_i$ is the gate insulator capacitance per unit area. In our TFTs, the estimated $C_i$ was approximately 13.8 nF/cm$^2$, based on a dielectric constant of 3.9 for SiO$_2$. The $\mu_{FE\text{-}i}$ obtained from the slope of these plots was 17.2, 22.9, and 18.0 cm$^2$/Vs for the as-fabricated, stored in vacuum, and exposure to air TFTs, respectively (Fig. 2(c)). For the TFTs stored in vacuum, the $V_{GS}$ dependence of the $r_{ch}$ correlates to metallic conduction with nearly constant values (Fig. 2(b)). The film conductivity and therefore the carrier density after storage significantly increased compared with the film before storage, although change in $\mu_{FE\text{-}i}$ was relatively small. This peculiar behavior in the intermediate carrier density region (~10$^{17}$ cm$^{-3}$) was typically observed in other amorphous oxide semiconductors and was explained using the percolation conduction model.[2,21,33] Because structural randomness leads to formation of further potential barriers,[34] an increase in the mobility derived from an increase in charge carrier density because of the creation of defective V$_O$ is limited.

Figure 2(d) summarizes the $R_c$ and transfer length difference ($\Delta L$) extracted from the horizontal axis of the intersection points for each linear fit.[35,36] $\Delta L$ is the distance from the source (or drain) electrode to actual edge of carrier injection (or correction) (Fig. 3). A positive value means the effective channel length ($L_{eff} = L_{ch} - \Delta L$) is shortened from the original electrode design.[37] For the TFTs stored in vacuum, $\Delta L$ increased with decreasing $R_c$. These changes correspond to a reduction in the $r_{ch}$ and an increase in $\mu_{FE-i}$ (Fig. 2(b) and (c)) because additional carriers are easily injected to the channel when the resistivity decreases, resulting in a reduction of the net barrier height of the contacts. Ueoka *et al.* claimed that $\Delta L$ is affected by the potential barrier height at the interface between oxide semiconductors and contact electrodes.[38] The experimental results revealed that the carrier density near the contact could be changed before/after the vacuum storage because of the desorption of excess oxygen molecules.[39] An X-ray absorption near-edge structure (performed at the BL27SU beamline of the SPring-8 synchrotron radiation facility) spectrum demonstrated that the coordination number of the Si atoms in the films was four under a vacuum of $10^2$ Pa. It was also confirmed that the Si-dopants tightly hold oxygen atoms around them. Thus the desorbed molecules under vacuum should be excess oxygen.

We chose $SiO_2$ as the oxygen binding dopant because of its high BDE, and prepared an ISO3 film at high $P_{O2}$ condition ($O_2$ concentration: 50 %) to adjust the $V_{ON}$ of the ISO3 film to be zero. Generally, in oxide thin-films deposited at higher $P_{O2}$, the carrier density monotonically decreases and the $V_{ON}$ of the fabricated TFTs shifts in the positive direction.[4,40] For realization of low-power-consumption TFTs, normally-off operation is strongly desired, so it is necessary to achieve $V_{ON} = 0$ V by adjusting $P_{O2}$ during sputtering. However, a vast amount of oxygen molecules are possibly incorporated into the film under a high $O_2$ concentration.[7,41,42] The molecules can act as additional electron traps because of the charge transfer from conduction paths of In ions and may suppress carrier generation.[6,43,44] Because the excess oxygen is non-bonded (or weakly-bonded) to In and Si atoms in the film, they easily diffuse out into ambient atmosphere as illustrated in the inset of Fig. 3, even when stored under a vacuum of around 10 Pa. Sallis *et al.* reported that excess oxygen could form deep subgap states acting as electron traps and could be removed by low-temperature annealing in a reducing atmosphere.[45] This would decrease the trap density in the oxide thin-film by desorbing excess oxygen molecules during the three months of vacuum storage. Thus, the $r_{ch}$ decreases and $V_O$-rich region ($\Delta L$) is extended. Subsequently, the TFT characteristics dramatically changed to metallic behavior. Recovering the *I-V* characteristics for the TFTs that were exposed to air is the opposite mechanism.

Based on the above discussion, to stabilize TFT characteristics under any environmental condition, charge carrier generation by oxygen desorption should be suppressed by increasing the oxygen binding dopant concentration. For reducing unstable excess oxygen and providing oxygen bonding in amorphous oxide semiconductors, to lower $P_{O2}$ during sputtering and to obtain the optimum dopant concentration with high BDE, a dopant that can be easily oxidized (with a low Gibbs free energy, $G_f$) is required.[46] SiO$_2$ has a high BDE and it also simultaneously possesses a low $G_f$ (−814 kJ/mol at 250 °C).[47] In this study, TFT characteristics using an In-Si-O film with a higher SiO$_2$ concentration (In$_2$O$_3$:SiO$_2$ = 90:10 by weight, hereafter denoted as ISO10) were demonstrated to obtain environmentally stable TFT operation.

Figure 4 shows *I-V* characteristics of ISO10 TFTs fabricated using the same scheme as the ISO3 TFTs. The $P_{O2}$ during ISO10 film deposition was reduced to 8.3 %. Also, ISO10 and ISO3 TFTs were stored in a same vacuum desiccator after fabrication for comparison. No $V_{ON}$ shift was observed in the ISO10 TFTs for all storage environments. $I_D$ was fully turned off for each channel length (Fig. 4(b)) even though no passivation layer was present. Superior transfer behavior was also observed after the TFT was exposed to air (Fig. 4(c)). The saturation $I_D$ in the output characteristics (Fig. 4(d)–(f)) was clearly observed in the high $V_{DS}$ regime. For the TFTs stored in vacuum, the *I-V* characteristics for the ISO10 films were different to those for the ISO3 films. This is because the presence of excess oxygen in the films was significantly reduced and the formation of $V_O$ was effectively controlled by adding more SiO$_2$. As an experimental proof of the suppression of oxygen diffusion out of the film, $R_c$ and $\Delta L$ did not change for each state and were almost constant as shown in Fig. 5(a). Also, the TLM measurements showed that all of the currents at low $V_{DS}$ fit the linear relation. As seen in the $r_{ch}$ and the $1/r_{ch}$ dependence on $V_{GS}$ (Fig. 5(b) and (c)), no difference was observed for the TFTs in all storage environments. These results indicate that the TFTs become environmentally stable when using ISO10 films sputtered with low $P_{O2}$ because of the suppression of excess oxygen. Compared with the ISO3 TFTs, the $\mu_{FE-i}$ of the ISO10 films decreased (Fig. 5(c)), however the TFT reliability improved. Additionally, the $\mu_{FE-i}$ could be enhanced by optimization of the oxygen binding dopant concentration and $P_{O2}$ during sputtering.

In summary, we compared the TFT characteristics using ISO films with low and high oxygen binding dopant concentrations in terms of out-diffusion of excess oxygen. In the case of ISO3 films deposited at high $P_{O2}$, the change in $R_c$ and $r_{ch}$ was very pronounced after being stored in vacuum and after exposure to air. This environmental instability depending on the storage conditions could be explained by desorption and absorption of excess oxygen based on the TLM analysis. By increasing the $SiO_2$ concentration (from 3 to 10 wt.%) and reducing $P_{O2}$ during sputtering (from 50.0 to 8.3 %), fluctuations in electrical properties caused by excess oxygen diffusing out of the films could be effectively suppressed. The resulting ISO10 TFTs became very stable in any storage environment. Even when the channel layer was exposed to ambient atmosphere in the long term, the TFTs still maintained their initial semiconducting properties. The results presented here are useful for the design of the concentration of oxygen binding dopant in a sputtering target and the $P_{O2}$ during sputtering.

**Acknowledgments**

The authors would like to thank Dr. X. Gao, Dr. M.-F. Lin, Dr. W. Ou-Yang (WPI-MANA), Dr. Y. Tamenori and Dr. A. Fujiwara (Japan Synchrotron Radiation Research Institute) for fruitful discussions, and K. Ohno (NIMS) for helpful experimental support. This work was partially supported by Grants-in-Aid for Scientific Research (26790051). SA acknowledges the JGC-S Scholarship Foundation. Synchrotron radiation experiments were carried out with the approval of the Japan Synchrotron Radiation Research Institute (Proposal 2014B1806).

**Figure captions**

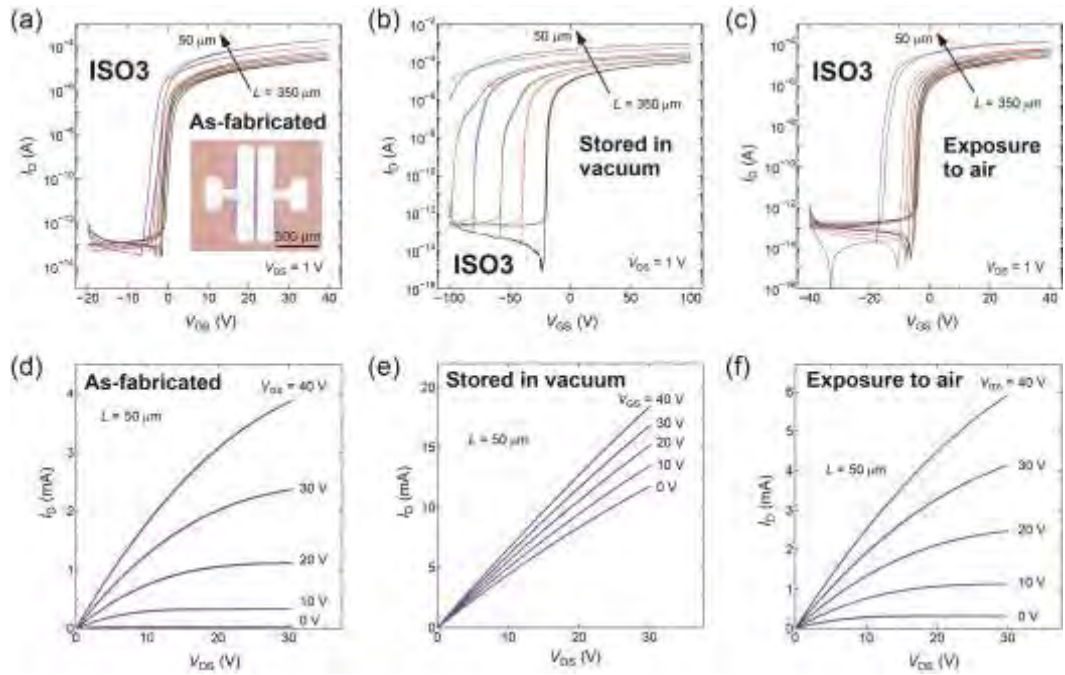

FIG. 1. *I-V* characteristics of thin-film transistors using ISO3 conditions. Transfer characteristics with different channel lengths (50−350 μm in 50-μm steps) for: (a) the as-fabricated TFTs; (b) the TFTs stored in vacuum (~10 Pa) for 3 months; (c) TFTs that were then exposed to air for 2 weeks. The inset in (a) is the optical microscope image of a TFT with $L/W$ = 50/1000 μm. (d–f) Output characteristics for the represented TFTs with $L$ = 50 μm.

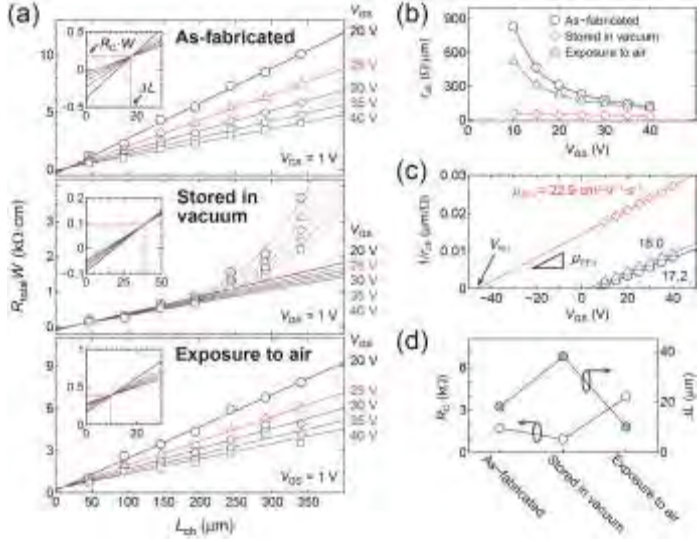

FIG. 2. Fitting of contact resistance ($R_c$) and channel resistivity ($r_{ch}$) for the ISO3 TFTs with channel lengths of 50−350 μm using the transfer line method. (a) The width-normalized total resistance ($R_{total}W$) extracted at drain voltage ($V_{DS}$) of 1 V as a function of channel length ($L_{ch}$), which were determined using optical microscope images. The inset shows the magnification of the intersection point. For the TFT stored in vacuum, the fitting lines were calculated using the plots less than $L_{ch}$ = 200 μm because of the nonlinear relation between $R_{total}W$ and $L_{ch}$. (b) The $r_{ch}$ and (c) reciprocal $r_{ch}$ as a function of gate voltage ($V_{GS}$). The values indicated in (c) are the intrinsic field-effect mobility ($\mu_{FE\text{-}i}$), which is extracted from the slope, whereas $V_{th\text{-}i}$ is the intrinsic threshold voltage obtained at the intersection of the slope and the horizontal axis. (d) The $R_c$ and transfer length difference ($\Delta L$) for the as-fabricated TFT, TFT stored in vacuum, and TFT that was exposed to air.

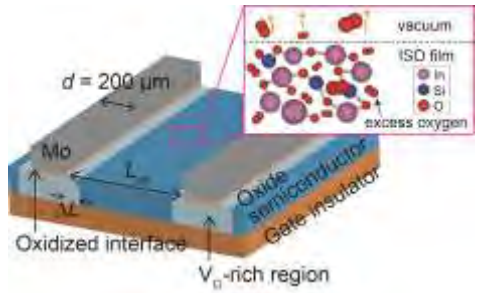

FIG. 3. Schematic diagram of the fabricated TFT with illustration of oxygen out-diffusion. For the as-fabricated TFT, an oxidized layer ($MoO_x$) is formed at the electrode/active channel interface by thermal annealing (Ref. 17). Oxygen molecules diffused to the interface during annealing treatment. The contact region becomes oxygen vacancy ($V_O$)-rich. The contact length (*d*) in all the TFTs was fixed at 200 μm. The inset shows a schematic illustration of the oxygen diffusion model. When the TFT was stored in vacuum, excess oxygen desorbed from the ISO film surface and inside, resulting in the decrease of the electron trap density of the film.

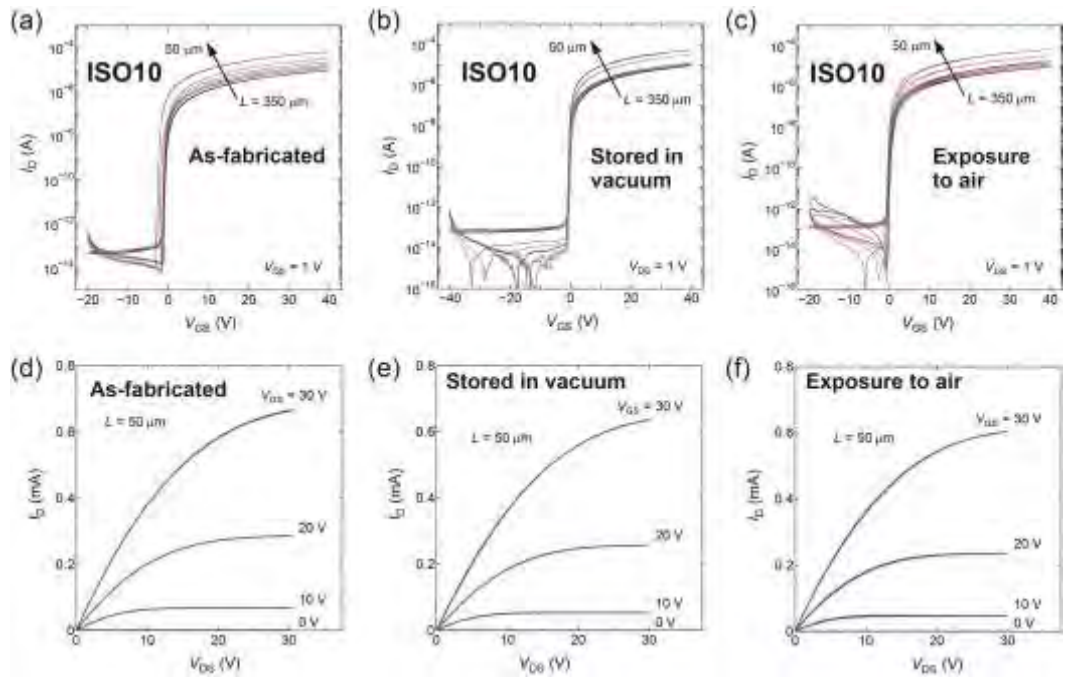

FIG. 4. (a–c) Transfer characteristics for the TFTs using ISO10 with different channel lengths (50–350 μm in 50-μm steps) for: (a) as-fabricated TFTs; (b) TFTs stored in vacuum; (c) TFTs that were then exposed to air. (d), (e), and (f) are the corresponding output characteristics for the transfer characteristics in (a), (b), and (c), respectively. The storage/exposure conditions (environment and period) are the same as the ISO3 TFTs.

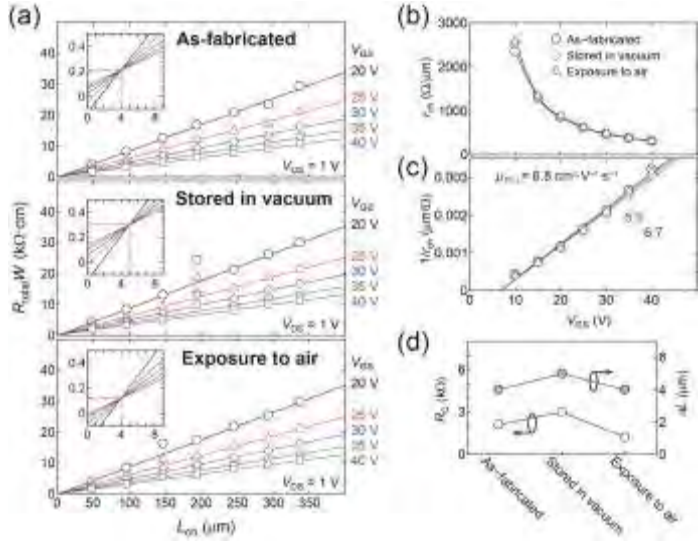

FIG. 5. (a) The $R_{total}W$ at $V_{DS}$ = 1 V as a function of $L_{ch}$ for the ISO10 TFTs. The inset shows the magnification of the intersection point. All the plots fit the linear relation. (b) The $r_{ch}$ and (c) reciprocal $r_{ch}$ as a function of $V_{GS}$. The values indicated in (c) are the $\mu_{FE\text{-}i}$. (d) The $R_c$ and $\Delta L$ (= $L_{ch} - L_{eff}$) for the as-fabricated, stored in vacuum, and exposure to air TFTs.